\documentclass[fleqn,11pt,twoside]{article}

\usepackage{amsthm,amsthm,amssymb, color, xcolor,epsfig, graphics, subfigure}

\usepackage{amsmath, graphicx, latexsym, lscape}

\makeatletter
\newcommand{\copyrightnote}[2]{{\renewcommand{\thefootnote}{}
 \footnotetext{\small\it
\begin{flushleft}
 \copyright \ #1   #2  
\end{flushleft}}}}

\newcommand{\Name}[1]{\begin{flushleft}
                       \LARGE \bf #1
                       \end{flushleft}\vspace{-3mm}}

\newcommand{\Author}[1]{\begin{flushleft}
                       \it #1 \end{flushleft}}

\newcommand{\Address}[1]{\begin{flushleft}
                       \it #1 \end{flushleft}}

\newcommand{\Date}[1]{\begin{flushleft}
                      \small  \it #1 \end{flushleft}}

\newcommand{\evenhead}{Author \ name}
\newcommand{\oddhead}{Article \ name}

\renewcommand{\@evenhead}{
\hspace*{-3pt}\raisebox{-15pt}[\headheight][0pt]{\vbox{\hbox to \textwidth
{\thepage \hfil \evenhead}\vskip4pt \hrule}}}
\renewcommand{\@oddhead}{
\hspace*{-3pt}\raisebox{-15pt}[\headheight][0pt]{\vbox{\hbox to \textwidth
{\oddhead \hfil \thepage}\vskip4pt\hrule}}}
\renewcommand{\@evenfoot}{}
\renewcommand{\@oddfoot}{}

\long\def\@makecaption#1#2{%
  \vskip\abovecaptionskip
  \sbox\@tempboxa{\small \textbf{#1.}\ \ #2}%
  \ifdim \wd\@tempboxa >\hsize
    {\small \textbf{#1.}\ \ #2}\par
  \else
    \global \@minipagefalse
    \hb@xt@\hsize{\hfil\box\@tempboxa\hfil}%
  \fi
  \vskip\belowcaptionskip}

\newcommand{\JNMPnumberwithin}[3][\arabic]{%
  \@ifundefined{c@#2}{\@nocounterr{#2}}{%
    \@ifundefined{c@#3}{\@nocnterr{#3}}{%
      \@addtoreset{#2}{#3}%
      \@xp\xdef\csname the#2\endcsname{%
        \@xp\@nx\csname the#3\endcsname .\@nx#1{#2}}}}%
}

\newcommand{\resetfootnoterule} {
  \renewcommand\footnoterule{%
  \kern-3\p@
  \hrule\@width.4\columnwidth
  \kern2.6\p@}
}

\renewcommand{\footnoterule}{}

\makeatother

\newcount\eqnumber
\def\neweq{{\rm{(\the\eqnumber)}}\global\advance\eqnumber by 1}
\def\eqdef#1{\eqno\xdef#1{\the\eqnumber}\neweq}
\def\newaeq{{\rm{(\the\eqnumber a)}}\global\advance\eqnumber by 1}
\def\eqdaf#1{\eqno\xdef#1{\the\eqnumber}\newaeq}
\def\eqdisp#1{\xdef#1{\the\eqnumber}\neweq}
\def\eqdasp#1{\xdef#1{\the\eqnumber}\newaeq}
\newcount\refnumber
\def\newref{{\the\refnumber}\global\advance\refnumber by 1}
\def\refdef#1{{\xdef#1{\the\refnumber}}\newref}
\def\bol{\scalebox{0.7}{$\bullet$}}
\usepackage{graphicx}
\usepackage{color}
\usepackage{hyperref}
\theoremstyle{definition}

\begin{document}

\renewcommand{\evenhead}{ {\LARGE\textcolor{blue!10!black!40!green}{{\sf \ \ \ ]ocnmp[}}}\strut\hfill 
R Willox, T Mase, A Ramani and B Grammaticos
}
\renewcommand{\oddhead}{ {\LARGE\textcolor{blue!10!black!40!green}{{\sf ]ocnmp[}}}\ \ \ \ \  
Singularities and growth of higher order discrete equations
}

%%%% Matter for the first page 
\thispagestyle{empty}
\newcommand{\FistPageHead}[3]{
\begin{flushleft}
\raisebox{8mm}[0pt][0pt]
{\footnotesize \sf
\parbox{150mm}{{Open Communications in Nonlinear Mathematical Physics}\ \ \ \ {\LARGE\textcolor{blue!10!black!40!green}{]ocnmp[}}
\quad Special Issue 2, 2024\ \  pp
#2\hfill {\sc #3}}}\vspace{-13mm}
\end{flushleft}}

\FistPageHead{1}{\pageref{firstpage}--\pageref{lastpage}}{ \ \ }

\strut\hfill

\strut\hfill

\copyrightnote{The author(s). Distributed under a Creative Commons Attribution 4.0 International License}

\begin{center}

{\bf {\large Proceedings of the OCNMP-2024 Conference:\\ 

\smallskip

Bad Ems, 23-29 June 2024}}
\end{center}

\smallskip

\Name{Singularities and growth of higher order discrete equations}

\Author{Ralph Willox$^{\,1}$, Takafumi Mase$^{\,1}$, Alfred Ramani$^{\,2}$ and Basil Grammaticos$^{\,2}$}

\Address{$^{1}$ Graduate School of Mathematical Sciences, the University of Tokyo, 3-8-1 Komaba, Meguro-ku, 153-8914 Tokyo, Japan \\[2mm]
$^{2}$ IMNC, Universit\'e Paris VII \& XI, CNRS, UMR 8165, B\^at. 440, 91406 Orsay, France}

\Date{Received March 22, 2024; Accepted April 10, 2024}

\setcounter{equation}{0}

\begin{abstract}

\noindent 
We study the link between the degree growth of integrable birational mappings of order higher than two and their singularity structures. The higher order mappings we use in this study are all obtained by coupling mappings that are  integrable through spectral methods, typically belonging to the QRT family, to a variety of linearisable ones.
We show that by judiciously choosing these linearisable mappings, it is possible to obtain higher order mappings that exhibit the maximal degree growth compatible with integrability, i.e. for which the degree grows as a polynomial of order equal to the order of the mapping. In all the cases we analysed, we found that maximal degree growth was associated with the existence of an unconfining singularity pattern. Several cases with submaximal growth but which still possess unconfining singularity patterns are also presented. In many cases the exact degrees of the iterates of the mappings were obtained  by applying a method due to Halburd, based on the preimages of specific values that appear in the singularity patterns of the mapping, but we also present some examples where such a calculation appears to be impossible.
\end{abstract}

\label{firstpage}

%%%% The Article text starts here

\section{Introduction}

There is a deep link between the singularity structure and the growth properties of birational mappings. In fact, the slow growth which characterises {\sl integrable} discrete systems [\refdef\veselov] is intimately related to the property known as singularity confinement [\refdef\sincon]. Both phenomena share a common origin: they are due to the simplifications [\refdef\blending] that can occur when the iteration of a birational mapping, starting from some initial values, 
hits an indeterminate point of the mapping. 
This is best illustrated by an example.  Consider the QRT [\refdef\qrt] mapping 
$$x_{n+1}x_{n-1}=a\left(1-{1\over x_n}\right),\eqdef\zena$$
which is known [\refdef\detecting] to possess the (confining) singularity pattern 
$$\{1,0,\infty,\infty,0,1\},$$
as well as the length-7 cyclic pattern
$$ \{\bol,0,\infty,\infty,0, \bol,\infty,\bol,0,\infty\dots\},$$
where the dots $\bol$ indicate finite, non-zero values that involve the initial conditions. 
In order to compute the degree growth of the mapping we introduce initial conditions $x_0=r$ and $x_1=p/q$ and compute the degree $d_n$ in $p,q$ of the successive iterates. We find thus the sequence of values:  0, 1, 1, 2, 3, 4, 6, 7, 10, 12, 15, 18, 21, 25, 28, 33, 37, 42, 47, 52, 58, 63, 70, 76, 83, 90, 97$\cdots$, which corresponds to a quadratic growth in the initial condition $x_1$, and thus to a {\sl dynamical degree} equal to 1. The latter is defined as 
$$\lambda=\lim_{n\to\infty}d_n^{1/n},\eqdef\intercal$$
and is therefore equal to 1 if the degree $d_n$ grows polynomially. A mapping with a dynamical degree that is equal to 1 is said to be integrable.

To illustrate how the singularity structure is intimately linked to this degree growth we start from (\zena) and the same initial conditions, $x_0=r$ and $x_1=p/q$ and iterate the map without implementing any simplifications. We thus find
$$x_2=a{p-q\over pr}$$
$$x_3={qP_1\over p(p-q)}$$
$$x_4={prP_2\over qP_1(p-q)}$$
$$x_5={pP_3(p-q)\over pqrP_2P_1}$$
$$x_6={pqP_4P_1(p-q)\over p^2rP_3P_2(p-q)}$$
$$x_7={p^2qrP_5P_2P_1(p-q)^2\over p^2qP_4P_3P_1(p-q)^2}$$
where the $P_n$ are polynomials in $p,q,r$ of degree $n$ in $p,q$. 
Note that the degrees of the iterates $x_2, \cdots, x_7$ written in this way, without simplifications, are: 1, 2, 3, 5, 8, 13.

First, let us follow the singularity induced by $p=0$. We obtain for $x_n$ the sequence $r,0,\infty,\infty,0,s,\infty,t$, where $s,\infty,t$ are the values obtained from $x_5$, $x_6$ and $x_7$ at $p=0$ once $p$ is divided out.  We remark that this succession of values follows precisely the cyclic pattern of (\zena). Next, we study the singularity induced by $p=q$. We now find for the $x_n$ the successive values $r,1,0,\infty,\infty,0$ and once the $(p-q)$ factor is divided out in $x_6$ a precise calculation of this iterate leads to the value 1. Similarly, a precise calculation of $x_7$ for $p=q$ leads to the value $r$, i.e. we recover the initial value $x_0$ lost in the iterations through the singularity at $x_1=1$. This phenomenon is what is commonly referred to as singularity confinement. The (finite length) singularity pattern induced by $p/q=1$ is precisely the confining pattern given above, and it is clear it could only arise because of the indeterminate nature of the iterates at $x_6$ and $x_7$. If a mapping solely has such confining patterns, possibly accompanied by cyclic patterns as the one above induced by 0, we say that the mapping has the {\sl singularity confinement} property.

As mentioned earlier, an important consequence of the successive simplifications in the above iterations is a lowering of the degree at each iteration as of $x_5$. Simple inspection shows that the degrees of the first 7 iterates are indeed 0, 1, 1, 2, 3, 4, 6, 7. Computing one more iterate we find
$$x_8={p^4qP_6P_3P_2P_1^2(p-q)^3\over p^3q^2P_5P_4P_2P_1^2(p-q)^3}$$
and the simplification removes a factor $p^3qP_2P_1^2(p-q)^3$, which leads to a degree of 10.

Such massive simplifications often have as result that the dynamical degree of the mapping is equal to 1, and thus that the mapping is integrable. Exceptions do exist however, for which the simplifications are sufficient for the singularities of the mapping to confine but do not curb its degree growth sufficiently to have a dynamical degree that is equal to 1. 

Obtaining the degree growth of a birational mapping, relying solely on its singularity structure, is perfectly possible thanks to the ingenious method proposed by Halburd in [\refdef\halburd].  This method is based on the observation that the degree $d_n$ of the $n$-th iterate of the mapping,  $f_n$, is equal to the number of preimages of some value of that function if one regards  $f_n$ as a (complex valued) function of just the single variable $x_1$. However, instead of finding the preimages for just any arbitrary value, Halburd proposes to use the special values that appear in the singularity patterns of the mapping. Nonetheless, in order to be able to derive a precise formula for the degree growth of the mapping, one must not limit oneself to the confining singularity patterns but  also take into account all contributions stemming from the cyclic patterns (even if, strictly speaking, some of these might not involve any singular points of the mapping). We illustrate this in the case of (\zena).
We denote by $Z_n$ the spontaneous appearance of the value 1 at some step $n$. But, as is clear from its confining singularity pattern, a value of 1 can also be the result of a spontaneous appearance of 1 five steps before. Thus the degree $d_n$ obtained from the preimage of 1 is given by 
$$d_n(1)=Z_n+Z_{n-5}.\eqdef\zdyo$$
The same degree is obtained when we count the number of occurrences of  0 and/or $\infty$ in the singularity patterns. Neglecting the contributions of the cyclic pattern, as one does in what we have called the ``express'' method [\refdef\express], would simply lead to the conclusion that the dynamical degree for this mapping is equal to one, something which is of course to be expected given the integrable character of (\zena). 
However, for a precise calculation of the degree, it is imperative that we consider the contribution of the cyclic pattern. We thus have
$$d_n(0)=Z_{n-1}+Z_{n-4}+f_7(n),\eqdef\ztri$$
$$d_n(\infty)=Z_{n-2}+Z_{n-3}+g_7(n),\eqdef\ztes$$
where $f_7$ and $g_7$ are periodic functions of $n$ of period 7: $f_7$ is equal to one when $n$ mod 7 is equal to 1,3,4,6 and 0 otherwise, and $g_7$ equal to 0 if $n=7m$ and equal to 1 otherwise.

Since the degree computed from (\zdyo) with either of the equations (\ztri) or (\ztes) is necessarily the same, we can establish an inhomogeneous linear equation for $Z_n$ involving $f_7$ and $g_7$ and solve it (assuming $Z_n=0$ for $n\le 0$) to find an explicit expression for $d_n$:
$$d_n=(n^2+4+\psi_7(n))/7.\eqdef\zpen$$
Here the periodic function $\psi_7(n)$ is obtained by the periodic repetition of the string $[-4,2,-1,1,1,-1,2]$, which is in perfect agreement with the degrees computed above.

The literature abounds with studies of the singularity structure of second-order birational mappings and there exist several methods for determining the precise growth of such a mapping [\refdef\early]. The case of higher-order mappings, despite not being a complete {\sl terra incognita} [\refdef\bessis,\refdef\yahagi,\refdef\capel,\refdef\redux,\refdef\lyness,\refdef\stef,\refdef\gub], has so far drawn less attention. 
The construction of integrable higher-order mappings starting from an integrable second-order one by coupling it to linearisable mappings was introduced by two of the authors, in collaboration with S. Lafortune in [\refdef\thrid], where it was called the {\sl Gambier approach}. This name refers directly to the construction of the Gambier mapping [\refdef\gambier], which consists of two homographic mappings coupled in cascade. An important point must be stressed here. It is well known that second-order linearisable mappings may exhibit non-confined singularities [\refdef\tremblay] and as we shall see throughout this paper, coupling an integrable mapping with confined singularities to one or more linearisable ones in order to construct higher-order mappings may also result in mappings with non-confined singularities, be they anticonfined [\refdef\anticonf] or genuinely unconfined [\refdef\unconf]. 

Most of the explicit examples we shall present in this paper are of third and fourth order. However, we believe that our conclusion that an analysis of the singularity structure of a mapping is indispensable if one wants to understand the structures that underlie the integrability of a higher order mapping, is in fact applicable to birational mappings of any order. As we shall see, at higher orders, whereas a mere study of the degree growth of a mapping does not give any information beyond the integrability or nonintegrability of the mapping, combining it with singularity analysis may actually provide insight into the way such a mapping might be integrated.

\section{Some exact results} 

Diller and Favre [\refdef\favre] have established the following classification of the different possible degree growths for second-order birational autonomous mappings, in terms of the properties of their singularities.
\smallskip

-- the degree growth can be bounded, which is of course the case for periodic mappings, but if such a mapping is non-periodic it is known to be birationally equivalent to a projective mapping on ${\bf P}^2$ [\refdef\deserti]. The singularity confinement property is irrelevant for such mappings.\medskip

-- the degree can grow linearly with $n$, in which case the mapping preserves a rational fibration over ${\bf P}^1$ and is de facto linearizable. Such mappings however do not possess the singularity confinement property: they necessarily possess non-confined singularities.\medskip

-- the degree can grow quadratically with $n$, in which case the mapping always has the confinement property and, moreover, preserves an elliptic fibration over ${\bf P}^1$.\medskip

In all of the above cases, the dynamical degree will be equal to 1. Such mappings are very special and generic second-order birational mappings will of course have exponential degree growth (i.e., they have a dynamical degree that is greater than 1). However, this does not necessarily preclude them from possessing the singularity confinement property (as exemplified by the example in [\refdef\hiv]).

A similar classification was obtained by one of the authors for non-autonomous second-order birational mappings, modulo certain general conditions on the type of non-autonomous behaviour that can be tolerated in the mapping [\refdef\masetwo].

For higher-order birational mappings, there is no established classification for their degree growth. There is however an intriguing conjecture formulated by Silverman in [\refdef\silver], which can be roughly summarized as stating that for a birational mapping $\varphi$ on ${\bf C}^N$ with dynamical degree $\lambda_\varphi$, the quantity ${\rm deg}(\varphi^n)\,  \lambda_\varphi^{-n}\, $ grows at most polynomially with $n$, for a polynomial with degree at most $N$. This conjecture has been shown to hold for monomial mappings for general $N$ (and holds for integrable mappings, i.e. mappings for which $\lambda=1$, when $N=2$), but as far as the authors are aware of it is still an open problem in all other cases. An obvious consequence of the conjecture is of course that the maximal growth for an integrable mapping would be $d_n\sim n^N$.

Another important result is that obtained in [\refdef\maseone] where it was shown that the degree of any mapping (possibly of order greater than 2) that can be obtained as a direct reduction of the A-or B-type discrete KP equations, can only grow as $n^\ell$ where $\ell$ is $0, 1$ or $2$. \hfill\break
A similar result is expected to hold for the C-and D-type discrete KP equations studied by Schief et al. [\refdef\schiefone,\refdef\schieftwo]. This result strongly suggests that mappings that arise `naturally' in the theory of discrete integrable systems, i.e. those which inherit some of the extremely rich geometric and algebraic structures that are present in the discrete KP equations, will never have degree growth faster than $n^2$ and any integrable mapping that does exhibit faster polynomial growth must therefore lose certain of the aforementioned properties. In this paper, we would like to argue that the property that will be lost in such cases is singularity confinement. 

Another important result in this respect, due to two of the present authors and collaborators, concerns the $M$-th order Gambier mapping. The latter has the form
$$x^{(1)}_{n+1}={a^{(1)}x^{(1)} +b^{(1)}\over c^{(1)}x^{(1)}+d^{(1)}},$$
$$x^{(i)}_{n+1}={(e^{(i)}x^{(i-1)}+f^{(i)})x^{(i)} +(g^{(i)}x^{(i-1)}+h^{(i)})\over
(j^{(i)}x^{(i-1)}+k^{(i)})x^{(i)}+(l^{(i)}x^{(i-1)}+m^{(i)})},\quad i=2, \dots, N. \eqdef\gamb$$
i.e., it consists of a cascade of homographic equations where the solution of each one enters linearly into the coefficients of the next one. As has been shown in [\refdef\steph], generically an $N$-th order Gambier mapping exhibits a degree growth  $d_n\sim n^{N-1}$. {Moreover, it is known that any mapping on ${\bf P}^1\times {\bf P}^1$ that has linear growth can, after some transformations, be expressed as the coupling of two homographies (see e.g. p.139 in [\refdef\bpv] and the proof of Lemma 4.2 in [\favre]).}

Combining these results we arrive at the conjecture that for integrable mappings on ${\bf C}^N$ the maximal degree growth will be $d_n \sim n^N$ and that this maximal growth can only be achieved by coupling at least one second-order mapping that exhibits quadratic growth with a sufficient number of linear (or linearizable) mappings.  

This will be the strategy adopted in the sections that follow, but before moving on to the main body of the paper let us first give a simple example on which the problems that we wish to address can be easily explained.

\bigskip
{\sl An interlude}
\medskip

An illustrative example of such a coupling of an integrable map with quadratic growth to a linearizable one, is that of  a discrete Painlev\'e equation considered in conjunction with the defining equation for the functions through which the independent variable enters its coefficients. Consider the discrete Painlev\'e equation
$$x_{n+1}x_{n-1}={q_n\over x_n}+{1\over x_n^2},\eqdef\sena$$
where $q_n=q_0\lambda^n$. The equation satisfied by $q_n$ is simply $q_{n+1}q_{n-1}=q_n^2$, which is obviously linearisable. Given the form of (\sena) we can eliminate $q_n$ and obtain for $x_n$ the fourth-order autonomous mapping
$$x_n^2(x_{n+2}x_{n+1}^2x_n-1)(x_{n-2}x_{n-1}^2x_n-1)-x_{n+1}x_{n-1}(x_{n+1}x_{n-1}x_n^2-1)^2=0.\eqdef\sdyo$$
\textcolor{black}{This mapping is equivalent (in the gauge $y_n=q_n x_n$) to the generalized Y-system for the Somos-4 recurrence (equation (2.12) in [\refdef{\andyrei}]).}
The degree growth of the fourth order mapping defined by (\sdyo) is readily obtained: we find the sequence $d_n$=0, 0, 0, 1, 3, 7, 14, 24, 38, 57, 81, 111, $\cdots$ which shows cubic growth. 
This result is of course easy to understand. The equation for $q_n$ considered by itself has a solution, $q_0^{1-n} q_1^n$, with linear degree growth. This growth is combined with the quadratic one of a QRT mapping, i.e. the autonomous limit of (\sena), leading to an overall cubic growth for the solutions of the coupled system (\sdyo). \textcolor{black}{Note that a similar cubic growth has been observed [\refdef\andyposter] for the tau functions for the discrete Painlev\'e equation (\sena), in terms of the parameter $\lambda$ in $q_n=q_0\lambda^n$ (the corresponding degree sequence for the case $\lambda=2$ is listed in the On-Line Encyclopedia of Integer Sequences as sequence A095708 [\refdef\intseq]).}

On the other hand, if we consider an additive discrete Painlev\'e equation, 
$$x_{n+1}+x_{n-1}={z_n\over x_n}+{1\over x_n^2},\eqdef\senadd$$
in which the independent variable enters through $z_n=\alpha n+\beta$, we do not expect an increase in the degree growth beyond the quadratic.
Indeed, the function $z_n$ now obeys the linear equation $z_{n+1}-2z_n+z_{n-1}=0$ the solution of which exhibits zero growth and adds nothing to the quadratic growth of the QRT mapping. The resulting mapping
$$x_{n+2}x_{n+1}-x_{n+1}x_{n}-x_{n}x_{n-1}+x_{n-1}x_{n-2}-{1\over x_{n+1}}+{2\over x_{n}}-{1\over x_{n-1}}=0,\eqdef\sdyodd$$
therefore also has quadratic growth: $d_n=0, 0, 0, 1, 2, 5, 8, 13, 18, 25, 32, 41, 50, 61, 72, 85, \cdots$. 

It is worthwhile to take a closer look at the difference between these two couplings at the level of their singularity structures. For the case of equation (\sdyodd), we find that starting from generic $x_0, x_1, x_2$ and $x_3=0$, we have a confined singularity with pattern $\{0,\infty^2,0\}$, which is exactly the same pattern as for the second order mapping (\senadd). (The meaning of the exponents in the pattern above is the following: if we introduce a small parameter $\epsilon$ in lieu of 0 then $0^k$ and $\infty^k$ stand for terms proportional to $\epsilon^k$ and $\epsilon^{-k}$).
 Similarly, for generic $x_0, x_1, x_2$ and $x_3=\infty$ we find that this initial sequence is part of an anticonfining pattern of the form $\{\cdots,\bol,\infty,\bol,\infty,\bol,\bol,\bol,\infty,\bol,\infty,\bol,\cdots\}$ which replaces the (nonsingular) cyclic pattern $\{\bol,\infty\}$ that we have for (\senadd) (see [\anticonf] for more details on anticonfined singularities). Note that there are many other singularity patterns for the fourth order mapping (\sdyodd), such as $\{\cdots,\bol,\infty,\bol,\infty,\bol,\bol,0,\bol,\infty,\bol,\infty,\bol,\cdots\}$, that start from a co-dimension 1 initial condition. There are of course many more that start from initial conditions with greater co-dimensions, however, in the following we shall only treat $N-$point mappings for which we will only consider singularity patterns that arise from co-dimension 1 initial conditions in which a non-generic value is specified for the initial condition with the `highest' index, i.e. $x_{N-1}$ if the first initial condition is given at $x_0$. In all cases that we consider, these patterns turn out to be sufficient to analyse the behaviour of the mappings.

Coming back to the difference between (\sdyodd) and (\sdyo), if we analyse the singularities that arise from special values for $x_3$ for the mapping (\sdyo) we find that it also inherits the confining pattern $\{0,\infty^2,0\}$ from (\sena), just as (\sdyodd), but that the original cyclic pattern of (\sena) is now replaced by an unconfining one which exhibits linear growth in the mutiplicities of the singular values: $\{\bol,\bol,\bol,\infty,\infty,\infty,\infty^2,\infty^2,\infty^2,\infty^3,\cdots\}$. This unconfining pattern has its origins in the anticonfining singularity pattern $\{\cdots,0^3,0^2,0^1,\bol,\infty^1,$ $\infty^2,\infty^3,\infty^4,\cdots\}$ that exists for the equation $q_{n+1}q_{n-1}=q_n^2$ through the relation $q_n=(x_{n+1}x_n x_{n-1} -1/x_n)$, which obviously links the linear growth in the above anticonfining pattern to that in the unconfining one for (\sdyo). On the other hand, the anticonfining pattern $\{\cdots,\infty,\infty,\bol,\infty,\infty,\cdots\}$ for the linear relation $z_{n+1}-2z_n+z_{n-1}=0$ for $z_n = (x_{n+1}+x_{n-1}) x_n-1/x_n$ in the additive equation (\senadd), clearly cannot produce such an unconfining pattern for the mapping (\sdyodd). Hence, though both mappings are integrable, due to these fundamentally different couplings the singularity structures of the mappings (\sdyo) and (\sdyodd) are very different, the unconfined singularity structure of the former leading to a cubic degree growth which is unattainable for the latter.

\section{Coupling QRT mappings to linearisable systems} 

As a first step in the study of the growth properties and singularity structures of higher-order mappings, we will concentrate on mappings of orders 3 and 4, which we are going to construct by coupling a QRT mapping to some linearisable equation. 

In fact, the simplest case we shall consider is when the latter is just a homographic mapping (although second-order 
linearisable equations will also be considered). As was the case in the example at the end of section 2., we will also restrict ourselves to couplings that allow us to write the resulting system explicitly as a 3rd or 4th order mapping. This has as a consequence that the variable of the initial QRT mapping must appear linearly in the coefficients of the homographic equation. In order to illustrate this we present a coupled mapping which was first studied in [\thrid]. 

Let us start with the QRT mapping
$$x_{n+1}+x_{n-1}={a\over x_n}+{1\over x_n^2},\eqdef\vena$$
which has the confining singularity pattern $\{0,\infty^2,0\}$, as well as the cyclic pattern $\{\bol,\infty,\bol,$ $\infty,\cdots\}$. 

If we wish the two patterns for (\vena) to appear in the coupled equation we can convince ourselves 
that the coupling should have the form $y_{n+1}=H(x_ny_n)$ where $H$ is a homography of its argument. In order to make the analysis that follows somewhat easier, we separate the general coupling into two subcases,
$${\rm A:}\qquad y_{n+1}=y_nx_n,\eqdef\vdyo$$
and,
$${\rm B:}\qquad y_{n+1}=y_nx_n+1,\eqdef\vtri$$
which cover the general case, up to a homography on $y_n$ and scaling of $x_n$. Note that these forms of the coupling $y_{n+1}=H(x_ny_n)$ favour singularities at 0 and $\infty$. There exist mappings where the singularity patterns involve more than two values. In this case, the coupling will map these extra values to values for $y$ which will have nothing special. Still, the analysis can proceed in exactly the same way.

The study of the growth of the resulting third-order mappings is straightforward. 
In the case  of coupling A we obtain the mapping
$$y_{n+1}=y_{n-1}\left(a+{y_{n-1}\over y_n}\right)-{y_ny_{n-1}\over y_{n-2}},\eqdef\xdyo$$
for which we find quadratic growth: the degree sequence is $d_n$= 0, 0, 1, 2, 4
, 6, 9, 12, 16, 20, 25, 30, 36, 42, 49, 56, 64, $\cdots$. In the case  of coupling B, we find the mapping
$$y_{n+1}=1+{y_{n}y_{n-1}(a(1-y_n)-y_{n-1})\over (1-y_n)^2}+{y_n(1+y_{n-1})\over y_{n-2}},\eqdef\xtri$$
and the degree sequence $d_n$= 0, 0, 1, 3, 7, 13, 22, 34, 50, 70, 95, 125, 161, 203, 252, 308, $\cdots$, clearly a cubic growth as the third difference of the terms in the sequence leads to a periodically repeating pattern $(0,1)$. How can we explain these two different growths? The answer can be found in their different singularity structure.
In the case of coupling A the singularity structure is simple. We have just
$$ \{ 0,\infty\}\quad {\rm (confined)}$$
$$ \{ \cdots,0^2,0,0,z_0,z_1,\infty,\infty,\infty^2,\infty^2,\infty^3,\infty^3,\cdots\}\quad {\rm (anticonfined)}$$ 
where the anticonfining pattern exhibits linear growth in its exponents.

The case of coupling B however has the following singularities, including an unconfined one:
$$ \{ 0,1\}\quad {\rm (confined)}$$
$$ \{ 1,\infty^2,\infty,\infty,\infty,\cdots\}\quad {\rm (unconfined)}$$
$$ \{ \cdots,\bol,0,\bol,0,y_0,y_1,\infty,\infty,\infty^2,\infty^2,\infty^3,\infty^3,\infty^4,\infty^4,\cdots\}\quad {\rm (anticonfined)}.$$
Applying Halburd's method to the patterns obtained from coupling B, and denoting respectively by $Z_n$ and $M_n$ the spontaneous appearance of the values 0 and 1 at some step, the degree $d_n$ obtained from the preimage of 0 is just
$$d_n=Z_n,\eqdaf\vtes$$
for $n\ge 0$ and that obtained from the preimage of $1$ has two contributions, from the confining and the unconfining patterns:
$$d_n=Z_{n-1}+M_n.\eqno(\vtes b)$$
The contribution of $\infty$ comes from the unconfining pattern as well as from the anticonfining one. 
For $n\ge 0$ we have
$$d_n=\left[n\over2\right]+2M_{n-1}+M_{n-2}+M_{n-3}+\cdots,\eqno(\vtes c)$$
where $[p]$ stands for the integer part of a real number $p$. 
Note that this expression is {\it not} valid for $n<0$. Moreover, since there are never any negative contributions, the fact that $d_0$ and $d_1$ vanish means that  $M_n$ must vanish for all $n\le 1$. 
From  (\vtes a,b) we find $M_n=d_n-d_{n-1}$ 
for $n\ge 1$ and substituting this expression into (\vtes c) therefore only yields a finite number of terms. Combining with (\vtes a), we obtain for the degree the equation (for $n\ge 1$)
$$d_{n+1}-2d_n+d_{n-1}=\left[n+1\over2\right]={2n+1-(-1)^n\over4}.\eqdef\vpen$$
Its solution, with initial conditions $d_0=0, d_1=0$ is 
$$d_n={n^3\over12}+{n^2\over8}-{n\over12}-{1\over16}(1-(-1)^n),\eqdef\vhex$$
which reproduces exactly the degree sequence we computed for this coupling (and where we readily see where the periodic term in the third difference comes from). For further reference we shall denote this degree as $d_n^{\{19\}}$.

The calculations in the case of coupling A proceed along the same lines. The degree $d_n$ obtained from the preimage of 0, (for $n\ge 0$ so we can ignore the contribution from the anticonfining pattern), is again
$$d_n=Z_n,\eqdaf\vhep$$
while the contribution of the $\infty$ coming from the anticonfining pattern is, also for $n\ge 0$ 
$$d_n=Z_{n-1}+\left[n\over2\right].\eqno(\vhep b)$$
The equation for the degree becomes now (for $n\ge 1$)
$$d_n-d_{n-1}={2n+1-(-1)^n\over4},\eqdef\voct$$
with solution, for initial conditions $d_0=0, d_1=0$
$$d_n={n^2\over4}-{1\over8}(1-(-1)^n).\eqdef\venn$$
Again the expression of $d_n$ reproduces exactly the degree sequence we have found. Note also that this expression coincides exactly with the difference $d_n^{\{19\}}-d_{n-1}^{\{19\}}$.

Next, we introduce one further coupling, resulting in a mapping of order 4. Clearly, we can combine the two couplings A and B in four different ways. The coupling through two successive A-type couplings, $y_{n+1}=y_nx_n$ and  $z_{n+1}=z_ny_n$ leads to the following equation:
$$z_{n+1}={z_{n-1}^3+az_{n}z_{n-1}z_{n-2}-z_{n}^2z_{n-3}\over z_{n-2}^2}.\eqdef\newzena$$
The confining pattern we had for the single A coupling has now disappeared and the anticonfining pattern which remains now exhibits quadratic growth in the exponents:
$$ \{ \cdots,\infty^4,\infty^2,\infty, z_0, z_1,z_2, \infty,\infty^2,\infty^4,\infty^6,\infty^9,\infty^{12},\cdots\}.$$
The growth in the exponents we see in the pattern coincides exactly with the degree growth of the mapping, $d_n$= 0, 0, 0, 1, 2, 4, 6, 9, 12, 16, $\cdots$, and is precisely the same we obtained for the third order mapping (\xdyo) resulting from a single A coupling.

Coupling first with A and then with B, $y_{n+1}=y_nx_n$ and  $z_{n+1}=z_ny_n+1$, we obtain  the mapping
$$z_{n+1}=1+{z_{n}(z_{n-1}-1) \Big(z_{n-1}(z_{n-2}-1)\big(az_{n-2}(z_{n}-1)+z_{n-1}(z_{n-1}-1)\big)-z_{n-3}z_{n-2}(z_n-1)^2\Big)\over (z_{n}-1)z_{n-1}(z_{n-2}-1)z_{n-2}^2}.\eqdef\newzdyo$$
The sequence of degrees for this mapping, $d_n$= 0, 0, 0, 1, 3, 7, 13, 22, 34, 50, $\cdots$ shows cubic growth and is, in fact, the very same sequence we found at order three for a single coupling of type B. The singularity structure for the variable of the fourth-order equation (\newzdyo) is actually quite similar to that for the mapping (\xtri) obtained from a single B coupling, albeit now with a quadratic growth of the exponents in the anticonfining pattern:
$$ \{ 0,1\}\quad {\rm (confined)}$$
$$ \{ 1,\infty,\infty,\infty,\infty,\cdots\}\quad {\rm (unconfined)}$$
$$ \{ \cdots,\infty^4,\infty^2,\infty, z_0, z_1,z_2, \infty,\infty^2,\infty^4,\infty^6,\infty^9,\infty^{12},\cdots\}\quad {\rm (anticonfined)}.$$
With the same notations as above we have that the degree $d_n$ obtained from the preimage of 0 is just
$$d_n=Z_n,\eqdaf\tena$$
 for $n\ge 0$ because the fact that $d_0=0$ precludes any contribution of possible unconfining patterns extending all the way to $-\infty$ in $n$ and ending with $0$. As before, the degree from the preimage of $1$ has two contributions, from both the confining and the unconfining patterns:
$$d_n=Z_{n-1}+M_n.\eqno(\tena b)$$
Because $d_0, d_1$ and $d_2$ are zero, $M_n$ must vanish for all $n\le 2$. The contribution of $\infty$ comes from the unconfining pattern with a source term coming from the anticonfining one. We find, for $n\ge 0$
$$d_n={n^2\over4}-{n\over2}+{1\over8}(1-(-1)^n)+M_{n-1}+M_{n-2}+M_{n-3}+\cdots.\eqno(\tena c)$$
From (\tena ab), for $n\ge 1$, we have $M_n=d_n-d_{n-1}$ for $n\ge 1$ and because $d_0=0$ the sum of the $M_k$ in (\tena c) is just $d_{n-1}$. So
$$d_n-d_{n-1}={n^2\over4}-{n\over2}+{1\over8}(1-(-1)^n),\eqdef\tdyo$$
as mentioned before, the integration of which leads precisely to the cubic sequence of degrees given in expression (\vhex): $d_n=d_{n-1}^{\{19\}}$ for $n\ge1$.

Coupling first with B and then with A leads to the mapping
$$z_{n+1}={z_{n-3}z_n(z_{n-2}-z_{n-1})(z_{n-1}-z_n)^2+z_{n-1}\big(z_{n-2}^2(z_{n-1}-z_n)^2-az_{n-2}z_{n-1}z_n(z_{n-1}-z_n)+z_{n-1}^3z_n\big)\over z_{n-1}z_{n-2}^2(z_{n-1}-z_n)^2}.\eqdef\newztri$$
Implementing two successive B-type couplings, on the other hand, would lead to a mapping which is unfortunately too complicated to be explicitly given here. Both mappings however have the same unconfined singularity: starting from the unconfined singularity  $ \{ 1,\infty^2,\infty,\infty,\infty,\cdots\}$ for the variable $y$ we find for $z$ the singularity pattern $ \{ \bol,\infty^2,\infty^3,\infty^4,\infty^5,\cdots\}$, for both couplings. Both mappings also have the same anticonfining singularity pattern, $\{\cdots, \infty^3\infty^2,\infty^2,\infty,\infty,\bol,\bol,\bol,\infty,\infty^2,\infty^4,\infty^6,\infty^9,\infty^{12},$ $\infty^{16},\cdots\}$, which exhibits quadratic growth in the exponents.
Both couplings (B-B as well as B-A) actually lead to the same degree sequence as well, $d_n$= 0, 0, 0, 1, 4, 11, 24, 46, 80, 130, 200, 295, 420, $\cdots$, showing quartic degree growth ($n\geq1$) : 
$$d_n={n^4\over48}-{n^2\over12}+{1\over32}(1-(-1)^n).\eqdef\vdek$$
Note that the first difference of this quartic function is precisely the cubic one we obtained in (\vhex): we have
$d_n^{\{19\}}=d_{n+1}-d_n$, where $d_n^{\{19\}}-d_{n-1}^{\{19\}}$ is the quadratic function (\venn). 
Computing the growth in this case by Halburd's method would have been interesting, but unfortunately, the structure of the singularities of these fourth-order mappings is such that there are simply not enough equations to perform the calculations.

At this point, it is interesting to examine one more A coupling in the case A+A, which still gave a quadratic growth for case of the fourth order equation (\newzena) although the singularity structure had significantly changed compared to the case of a single A coupling. 
From the A+A+A coupling of (\vena) we obtain the mapping
$$z_{n+1}={z_{n}z_{n-3}\over z_{n-2}^2}\left({z_{n-1}^3z_{n-3}\over z_{n-2}^3}+az_n-{z_{n}^2z_{n-3}^2\over z_{n-1}^2z_{n-4}}\right),\eqdef\newztes$$
for which we find that its growth is no longer quadratic but now becomes cubic. 
It is easy to see where this cubic growth stems from. Besides the unconfined singularity $\{\bol, 0,0,0,\cdots\}$ this mapping also has an anticonfined one,
$$ \{ \cdots, 0,w_0,w_1,w_2, w_3, \infty,\infty^3,\infty^7,\infty^{13},\infty^{22},\infty^{34},\infty^{50},\infty^{70},\cdots\}\quad {\rm (anticonfined)},$$
on which a cubic growth of the exponents is clear. The overall degree growth of the mapping cannot be smaller than this and, when computed explicitly, turns out to be precisely that of the anticonfining pattern. It is given by (\vhex), shifted by two steps: $d_n=d_{n-2}^{\{19\}}$ for $n\ge2$.

Next, we turn to another QRT mapping (which is known to lead to a $q$-discrete Painlev\'e equation [\refdef\dps])
$$x_{n+1}x_{n-1}={x_n-a\over x_n-b}.\eqdef\ttri$$
It has two confining singularity patterns, $\{a,0,1/b,\infty,b\}$ and $\{b,\infty,1/b,0,a\}$, as well as a cyclic one $\{\bol,\infty,\bol,0,\cdots\}$. We couple it with a mapping of type B and obtain the third-order mapping
$$y_{n+1}=1+{y_{n-2} y_n\over1-y_{n-1}}{1-y_n+ay_{n-1}\over1-y_n+by_{n-1}}.\eqdef\xena$$
The confined singularities of (\ttri) induce an unconfined singularity $\{\bol, \infty,\infty,\infty,\cdots\}$ for (\xena), while the cyclic pattern of (\ttri) leads to the cyclic pattern $\{\bol,\bol,\infty,\infty\}$.
Computing explicitly the degree we obtain the sequence, $d_n$= 0, 0, 1, 2, 3, 5, 8, 12, 17, 24, 32, 42, 54, 68, 84,
103, 124, 148, $\cdots$, we can confirm that it has cubic growth and, moreover, obtain an explicit expression for the degree as a function of $n$ (valid for $n>1$)
$$d_n={n^3\over36}+{n^2\over24}-{n\over12}+{137\over144}-{(-1)^n\over16}+{1-j\over 27}j^n+{1-j^2\over27}j^{2n},\eqdef\ttes$$
where $j$ is a (complex) cubic root of unity. 
\bigskip

{\sl Coupling to second-order linearisable mappings}
\medskip

Up to now, we have considered only couplings of a QRT mapping to a first-order linearisable (and in practice, linear) equation. However, it is interesting to analyse the case where the coupling is to a second-order linearisable equation. Two such cases were already presented in section 2., but here we will examine two more cases, one where the coupling is through a projective mapping and one where the second-order mapping is of Gambier type. 

The general form of the projective second-order equation has been given, for instance, in [\refdef\karra] and [\refdef\again]. In this case, we shall consider a simplified form of it and use the coupling
$$z_{n+1}={x_n\over z_nz_{n-1}}.\eqdef\tpen$$
Using this coupling with (\vena) we obtain the fourth-order mapping
$$z_{n+1}z_{n}^3z_{n-1}^3z_{n-2}^2+z_{n}^2z_{n-1}^3z_{n-2}^3z_{n-3}+az_{n}z_{n-1}z_{n-2}+1=0.\eqdef\thex$$
Its singularity patterns are
$$ \{\bol,\bol,\bol,\infty,0,\infty\}\quad {\rm (cyclic)},$$
$$ \{ 0,\infty^3,0^3,\bol,\infty^3,0^3,\bol,\infty^3,0^3,\cdots\}\quad {\rm (unconfined)}.$$
Direct computation of the degree growth for this mapping leads to the sequence $d_n$= 0, 0, 0, 1, 3, 7, 12, 21, 33, 49, 69, 94, 123, $\cdots$, which exhibits cubic growth. Given the structure of the singularities, we can apply Halburd's method for the calculation of the degree. Following the same procedure as before we find that the degree obtained from the preimage of 0 is
$$d_n=Z_n+3(Z_{n-2}+Z_{n-5}+Z_{n-8}+\cdots)+\sum_{p=0}^{\left[n-4\over6\right]}\delta_{n-4-6p},\eqdaf\thep$$
while for $\infty$ we have  
$$d_n=3(Z_{n-1}+Z_{n-4}+Z_{n-7}+\cdots)+\sum_{p=0}^{\left[n-3\over6\right]}\delta_{n-3-6p}+\sum_{p=0}^{\left[n-5\over6\right]}\delta_{n-5-6p}.\eqno(\thep b)$$
The terms involving the $\delta_n$ correspond to the contributions from the anticonfining pattern. Notice that the series appearing in the two expressions are in fact finite sums, since $Z_k=0$ for $k\le2$ so as to have  $d_n=0$ for $n=0, 1, 2$. Combining the two equations we find
$$Z_n=d_n-d_{n-1}+\sum_{p=0}^{\left[n-6\over6\right]}\delta_{n-6-6p},\eqdef\toct$$
and substituting back into (\thep) we can obtain explicit expressions for $Z_n$ and $d_n$. We shall not enter into these details but just give the result: for $Z$ we find that $Z_n=n^2/4-n/2+(1-(-1)^n)/8-\sum_{p=0}\delta_{n-6-6p}$ and finally for the degree
$$d_n={n^3\over12}-{n^2\over8}-{n\over12}+{1\over16}(1-(-1)^n)-\left[n\over6\right],\eqdef\tenn$$
which reproduces precisely the degree sequence we have obtained above.

We turn now to a coupling of the mapping (\vena) with the Gambier-type mapping we introduced in [\gambier],
$$z_{n+1}-z_n+{1\over z_n}-{1\over z_{n-1}}=x_n.\eqdef\tdek$$
This leads to a fourth-order mapping of the form
$$z_{n+1}=z_n - z_{n-1} + z_{n-2} + {1+ {a (z_n - z_{n-1} + {1\over z_{n-1}} - {1\over z_{n-2}})}\over (z_n - z_{n-1} + {1\over z_{n-1}} - {1\over z_{n-2}})^2} - {1\over z_n} + {1\over z_{n-1}} - {1\over z_{n-2}} + {1\over z_{n-3}}.\eqdef\qena$$
The singularity patterns of (\qena) are
$$ \{ 0,\infty\}\quad {\rm (confined)}$$
$$\{\cdots, 0,0,\bol,\bol,0,0,z_0,z_1,z_2,\infty,\infty,\bol,\bol,\infty,\infty,\cdots\}\quad {\rm (anticonfined)}$$
$$ \{z_0, z_1, z_2, z_3=z_2-{1\over z_2}+{1\over z_1},\infty^2,\infty^2,\infty^2,\infty^2,\cdots\}\quad {\rm (unconfined)}$$
for generic values of $z_0, z_1,z_2$.
Calculating the degree growth directly on this mapping leads to the sequence $d_n$= 0, 0, 0, 1, 4, 10, 24, 51, 96, 164, 264, 405,$\cdots$, which exhibits quartic growth and can be represented by the expression
 $$d_n={1\over24}\big(n^4-4n^3+2n^2+16n+\psi_4(n)\big),\eqdef\qdyo$$
where $\psi_4(n)$ is a periodic function obtained by the repetition of the string $[0,-15,-24,-15]$. Note that the singularity structure for this mapping is such that Halburd's method cannot be applied directly to obtain its exact degree growth.
 
 \section{Coupling linearisable mappings to other linearisable ones} 
 
In section 2 we saw that an $N$th order Gambier mapping leads to a degree growth  $d_n\sim n^{N-1}$. Thus if we couple a second-order Gambier mapping, which has a linear degree growth, to a homographic mapping we expect the resulting mapping, in general, to have quadratic degree growth. Coupling two second-order Gambier mappings is expected to lead to a cubic degree growth and so on. However, as we shall see in the following, there exist departures from this maximal growth scenario depending on the details of the mappings involved.

As we have explained in previous publications (see for instance [\refdef\mimura]) there exist, at order two, three different families of linearisable mappings: projective ones, Gambier-type mappings and the family we have dubbed ``third-kind''. All the known examples of the latter belong to the QRT family at the autonomous limit, but, as is the case for all second-order linearisable systems, contain arbitrary functions of the independent variable. We shall start by studying the coupling of the third-kind mapping
$$x_{n+1}x_{n-1}=x_n^2-1,\eqdef\qtri$$
with the Gambier mapping (\tdek). We thus obtain a fourth-order mapping which is quite involved and will not be given here, but which has two singularity patterns, a confining one $ \{ 0,\infty\}$ and an anticonfining one $ \{\cdots, 0^2,0,\bol,\bol,\bol,\bol,\bol, \infty,\infty^2,\infty^3,\cdots\}$. The direct computation of the degree yields the following sequence $d_n$= 0, 0, 0, 1, 3, 7, 13, 21, 31, 43, 57, 73, 91, 111, 133, 157$\cdots$, which corresponds to a quadratic growth and not a cubic one. For $n\ge3$ the sequence of degrees can be represented by the expression $d_n=n^2-5n+7$. 

Another interesting coupling is of (\qtri) with twice the mapping A, which amounts to introducing
$$z_{n+2}={x_nz_{n+1}^2\over z_n}.\eqdef\qtes$$ 
We obtain thus the fourth-order mapping
$$z_{n+1}z_{n-3}z_{n-1}^6+z_{n-2}^2z_n^2(z_{n-1}^4-z_n^2z_{n-2}^2)=0.\eqdef\qpen$$
It has two singularity patterns:
$$\{z_0,z_1,z_2,z_3=\pm z_2^2/z_1,0,0^2,0^3,0^4,\cdots\}\quad {\rm (unconfined)}$$
$$ \{\cdots,\infty^{120},\infty^{84},\infty^{56},\infty^{35},\infty^{20},\infty^{10},\infty^4,\infty,z_0,z_1,z_2,0,0^2,0^2,\bol,\infty^5,\infty^{14},\infty^{28},\infty^{48}\cdots\}\quad {\rm (anticonfined)}.$$
Note that for the unconfining pattern, we start from generic $z_0,z_1,z_2$ but chose a special value for $z_3$ to enter the singularity.

The result is that the degree sequence obtained by direct calculation, namely $d_n$= 0, 0, 0, 1, 4, 10, 20, 40, 70, 112, 168, 240, 330, 440, $\cdots$, this time indeed corresponds to cubic growth,  as expected. For $n\geq6$ the degrees are represented by the expression
$$d_n={(n-1)(n-2)(n-3)\over3}.\eqdef\qhex$$

Next, we consider the coupling of the Gambier mapping
$$x_{n+1}-x_n+{1\over x_n}-{1\over x_{n-1}}=0,\eqdef\qhep$$
with a simple projective one
$$z_{n+1}={x_n\over z_nz_{n-1}}.\eqdef\qoct$$
The resulting fourth-order mapping is
$$z_{n+1}z_{n-3}z_{n-2}z_{n-1}^2z_n^2+z_{n-3}(1-z_n^2z_{n-1}^2z_{n-2}^2)-z_n=0.\eqdef\qenn$$
Two singularity patterns exist here as well:
$$\{0,\infty^2,0,0,\infty^2,0,0,\cdots\}\quad {\rm (unconfined)}$$
$$ \{\cdots,0,\bol,\bol,z_0,z_1,z_2,,\infty,\bol,\bol,\infty,\cdots\}\quad {\rm (anticonfined)}.$$
The direct calculation of the degree sequence however leads to $d_n$= 0, 0, 0, 1, 2, 4, 7, 10, 14, 19, 24, 30, 37, 47, 52,$\cdots$, and hence a quadratic growth which for $n\ge1$ can be represented by
$$d_n={n^2\over3}-n+{7\over 9}+{j^n+j^{2n}\over9},\eqdef\qdek$$
where $j$ is a complex cubic root of unity. Note that this growth is different from the cubic growth one would have obtained by coupling with a Gambier mapping instead.

The expressions for (\qhex) and (\qdek) can, in principle,  also be obtained from Halburd's method but the calculations are quite involved and we shall not go into these details.

Finally, we shall consider two different couplings of the Gambier mapping
$$(x_{n+1}+x_n)(x_n+x_{n-1})=a(x_n^2-1).\eqdef\uena$$
This mapping, although linearisable, has the special property of having confined singularities, with patterns $\{\pm1,\mp1\}$. As a consequence of the result by Diller-Favre (\favre) the degree of the successive iterates for such a mapping must be bounded, and we have in fact the sequence 0, 1, 2, 2,$\cdots$. Coupling (\uena) with another Gambier mapping, using (\tdek),  or with a projective mapping, using (\qoct), it turns out that while in the former case the degree growth is linear $d_n=0, 0, 0, 1, 3, 5, 7, 9, 11, \cdots$, in the second case it is in fact quadratic: $d_n=(n-3)^2$ (for $n\geq3$). The singularity structures that appear in both couplings turn out to be very different. For the fourth order mapping that results from coupling of (\uena) with the projective mapping (\qoct), and which has quadratic growth, we find two unconfining singularity patterns
$$\{0, 0^2, 0^3, 0^4, 0^5, 0^6,0^7,\cdots\}$$
$$\{\infty,\infty^3,\infty^6,\infty^{10},\infty^{15},\infty^{21},\infty^{28},\infty^{36},\cdots\},$$
the second one of which exhibits quadratic growth in the exponents that appear in it. For the mapping that is the result of a coupling to the Gambier mapping (\tdek) and which turned out to only have linear growth, however, we find that the resulting mapping still possesses one confining pattern, $\{0,\infty\}$ which is now accompanied by an unconfining one that shows no growth in the exponents that appear in it:
$$\{\infty,\infty,\infty,\cdots\}.$$
 
\section{Conclusions} 

In this paper we studied the singularity structures and growth properties of higher-order integrable mappings that are obtained through coupling various types of first and second order mappings.
We showed that, in such mappings, all four types of singularities, confined, cyclic, anticonfined and unconfined can appear and will play a role in determining the degree growth of the mapping. A particularly important finding is the existence of mappings with unconfined singularities, or with anticonfined singularities exhibiting growth in the associated singularity pattern, but which are still integrable mappings in that their dynamical degree is equal to 1. 

As discussed in Section 2, an integrable $N$th order mapping can have a degree growth $d_n\sim n^N$, and we have given several examples of couplings that achieve this maximal growth. For a fully linearisable mapping, i.e. fully linearisable by means of birational transformations, the degree growth is not faster than $d_n\sim n^{N-1}$. On the other hand, one does not expect a higher-order mapping integrable through spectral methods to have a growth faster than $n^2$ (which is, as explained, the maximal growth for mappings that are obtained as direct reductions of 3 dimensional lattice equations such as the discrete KP or BKP equations [\maseone]). Therefore it seems that in order to obtain a mapping with maximal growth, one must couple a mapping with quadratic growth with a linearisable one. In our paper, we illustrated this with mappings of orders three and four. In every case, it turned out that the singularity structure of the mapping involved one unconfining pattern. The case of the coupling of linearisable mappings was also considered resulting, in some cases, to the maximal possible growth but in some others in a submaximal one depending on the precise structure of the singularities of the system. The role played by couplings with Gambier mappings is particularly intriguing as the effect such a coupling has on the singularity structure and degree growth of the resulting mapping seems to depend greatly on the mapping it is coupled with.

Moreover, whenever an integrable mapping with unconfined singularities appears we consider this as an indication of the presence of a coupling to one or more linearisable equations. Finding the latter and uncoupling the mapping (which is more easily said than done) would result in an integrable confining mapping, including cases where the base equation is just a homography on ${\bf P}^1$.

We have also shown that in many cases one can use the singularity structure in order to exactly derive the degree growth using the method proposed by Halburd. However, this is not always possible. In some cases, the singularities are such that they do not allow one to establish a sufficient number of equations needed for the calculation of the number of preimages and thus of the degree $d_n$. This is something that was already observed in the case of second-order mappings and we found that this problem is present for higher-order systems as well. One way to address this difficulty is to introduce auxiliary variables which increase the number of equations to be satisfied by the preimages. In many cases, this turns out to be sufficient allowing the direct application of Halburd's method [\express].

\subsection*{Acknowledgements}

Two of the authors, RW and TM, would like to thank the Japan Society for the Promotion of Science (JSPS) for financial support through the KAKENHI grants 22H01130 and 18K03355 (RW), and 23K12996 and 18K123438 (TM).

\label{lastpage}

\begin{thebibliography}{99}

\bibitem{veselov} A.P. Veselov, {\sl Growth and integrability in the dynamics of mappings}, Commun. Math. Phys. 145 (1992) 181.

\bibitem{sincon} B. Grammaticos, A. Ramani and V. Papageorgiou, {\sl Do integrable mappings have the Painlev\'e property?}, Phys. Rev. Lett. 67 (1991) 1825.

\bibitem{blending} S. Lafortune, A. Ramani, B. Grammaticos, Y. Ohta and K.M. Tamizhmani, {\sl Blending two discrete integrability criteria: singularity confinement and algebraic entropy}, CRM proceedings Vol 29 (2001) p.299.	

\bibitem{qrt} G.R.W. Quispel, J.A.G. Roberts and C.J. Thompson, {\sl Integrable mappings and soliton equations II}, Physica D 34 (1989) 183.

\bibitem{detecting} B. Grammaticos, A. Ramani, R. Willox and T. Mase, {\sl Detecting discrete integrability: the singularity approach}, in Nonlinear Systems and Their Remarkable Mathematical Structures, N. Euler (ed), CRC Press, Boca Raton FL (2018) p 44.

\bibitem{halburd} R.G. Halburd, {\sl Elementary exact calculations of degree growth and entropy for discrete equations}, Proc. R. Soc. A 473 (2017) 20160831.

\bibitem{express} A. Ramani, B. Grammaticos, R. Willox and T. Mase, {\sl Calculating algebraic entropies: an express method}, J. Phys. A: Math. Theor. 50 (2017) 185203.

\bibitem{early} B. Grammaticos, A. Ramani, R. Willox, T. Mase and J. Satsuma, {\sl Singularity confinement and full-deautonomisation: A discrete integrability criterion}, Physica D 313 (2015) 11.

\bibitem{bessis} D. Bessis, C. Itzykson and J. B. Zuber, {\sl Quantum Field Theory Techniques in Graphical Enumeration}, Adv. in Math. 1 (1980) 109.

\bibitem{yahagi} R. Hirota, K. Kimura and H. Yahagi, {\sl How to find the conserved quantities of nonlinear discrete equations}, J. Phys. A: Math. Gen. 34 (2001) 10377.

\bibitem{capel} H.W. Capel and R. Sahadevan, {\sl A new family of four-dimensional symplectic and integrable mappings}, Physica A 289 (2001) 86.

\bibitem{redux} B. Grammaticos, A. Ramani, J. Satsuma, R. Willox and A.S. Carstea, {\sl Reductions of Integrable Lattices}, J. Nonlin. Math. Phys. 12, supp 1 (2005) 363.

\bibitem{lyness} B. Grammaticos, A. Ramani and T. Tamizhmani, {\sl Investigating the integrability of the Lyness mappings}, J. Phys. A: Math. Theor. 42 (2009) 454009.

\bibitem{stef} A.S.Carstea and T. Takenawa, {\sl Space of initial conditions and geometry of two 4-dimensional discrete Painlev\'e equations}, J. Phys. A: Math. Theor. 52 (2019) 275201.

\bibitem{gub}  G. Gubbiotti, N. Joshi, D.T. Tran and C.-M. Viallet, {\sl Bi-rational maps in four dimensions with two invariants}, J. Phys. A: Math. Theor. 53 (2020) 115201.

\bibitem{thrid} S. Lafortune,  B. Grammaticos and A. Ramani, {\sl Constructing integrable third-order systems: the Gambier approach}, Inv. Prob. 14 (1998) 287.

\bibitem{gambier}  B. Grammaticos, A. Ramani and S. Lafortune, {\sl The Gambier Mapping revisited},  Physica A 253 (1998) 260.

\bibitem{tremblay} A. Ramani, B. Grammaticos and S. Tremblay, {\sl Integrable systems without the Painlev\'e property}, J. Phys. A: Math. Gen. 33 (2000) 3045.

\bibitem{anticonf} T. Mase, R. Willox, B. Grammaticos and A. Ramani, {\sl Integrable mappings and the notion of anticonfinement}, J. Phys. A: Math. Theor. 51 (2018) 26520.

\bibitem{unconf} A. Ramani, B. Grammaticos, R. Willox, T. Mase and J. Satsuma, {\sl Calculating the algebraic entropy of mappings with unconfined singularities}, J. Integr. Sys. 3 (2018) xyy006.

\bibitem{favre} J. Diller and C. Favre, {\sl Dynamics of bimeromorphic maps of surfaces}, Amer. J. Math. 123 (2001) 1135.

\bibitem{deserti} J. Blanc and J. D\'eserti, {\sl Degree growth of birational maps of the plane}, Annali della Scuola Normale Superiore di Pisa Classe di Scienze 14 (2015) 507.

\bibitem{hiv} J. Hietarinta and C. Viallet, {\sl Singularity Confinement and Chaos in Discrete Systems}, Phys. Rev. Lett. 81, (1998) 325.

\bibitem{masetwo} T. Mase, {\sl Studies on spaces of initial conditions for non-autonomous mappings of the plane}, J. Integr. Sys. 3 (2018) xyy010.

\bibitem{silver} J. Silverman, {\sl Dynamical degree, arithmetic entropy, and canonical heights for dominant rational self-maps of projective space}, Ergodic Theory and Dynamical Systems 34 (2012) 647.

\bibitem{maseone} T. Mase, {\sl Investigation into the role of the Laurent property in integrability}, J. Math. Phys. 57 (2016) 022703.

\bibitem{schiefone} W. Schief, {\sl Lattice Geometry of the Discrete Darboux, KP, BKP and CKP Equations. Menelaus’ and Carnot’s Theorems}, J. Nonl. Math. Phys. 10 (2003) 194.

\bibitem{schieftwo} A.D. King and W. Schief, {\sl Bianchi Hypercubes and a Geometric Unification of the Hirota and Miwa Equations}, IMRN 16 (2015) 6842.

\bibitem{steph} S. Lafortune, B. Grammaticos and A. Ramani, {\sl Discrete and continuous linearizable equations}, Physica A 268 (1999) 129.

\bibitem{bpv} W. Barth, C. Peters and A. van de Ven, {\sl Compact Complex Surfaces}, Springer-Verlag, Berlin (1984).

\bibitem{andyrei} A.N.W. Hone and R. Inoue, {\sl Discrete Painlev\'e equations from Y-systems}, J. Phys. A: Math. Theor. 47 (2014 ) 474007.

\bibitem{andyposter} A.N.W. Hone, {\sl Algebraic curves, integer sequences and a discrete Painlev\'e transcendent}, preprint (2008) \url{https://arxiv.org/pdf/0807.2538.pdf}.

\bibitem{intseq} Sequence A095708 in the On-Line Encyclopedia of Integer Sequences:  \url{https://oeis.org/A095708}.

\bibitem{dps} A. Ramani, B. Grammaticos and J. Hietarinta, {\sl Discrete versions of the Painlev\'e equations}, Phys. Rev. Lett. 67 (1991) 1829.

\bibitem{karra} A. Ramani, B. Grammaticos and G. Karra, {\sl Linearizable mappings}, Physica A 180 (1992) 115.

\bibitem{again} A. Ramani, B. Grammaticos, K.M. Tamizhmani and S. Lafortune, {\sl Again, linearizable mappings}, Physica A 252 (1998) 138.

\bibitem{mimura} A. Ramani, B. Grammaticos, J. Satsuma and N. Mimura, {\sl Linearisable QRT mappings, }, J. Phys. A: Math. Theor. 44 (2011) 425201.

\end{thebibliography}
\end{document}